# Flexible terahertz opto-electronic frequency comb light source tunable over 3.5 THz

Dominik Theiner[1,2,*], Benedikt Limbacher[1,2], Michael Jaidl[1,2], Karl Unterrainer[1,2] and Juraj Darmo[1]

[1]TU Wien, Institut für Photonik, Gusshausstrasse 27-29, 1040 Vienna, Austria
[2]TU Wien, Zentrum für Mikro- und Nano-Strukturen, Gusshausstrasse 27-29, 1040 Vienna, Austria
*Corresponding author: dominik.theiner@tuwien.ac.at

**Abstract:** We demonstrate a terahertz (THz) frequency comb that is flexible in terms of its frequency range and the number and spacing of comb lines. We use a combination of near-infrared laser diodes, phase modulation and opto-electronic frequency conversion. The THz comb lines are characterized to be < 10 MHz by resolving the pressure-dependent collisional broadening of an ammonia molecule rotational mode.

The availability of a convenient light source is a key enabler for assessing the chemical compositions of objects and frequency combs (FCs) are ideal tools for this task [1,2]. The majority of environmentally and medically relevant molecules have their fundamental optically active rotational modes in the terahertz region (0.3-10 THz), where the generation of FCs is still very challenging. Only recently, THz combs were demonstrated from semiconductor Quantum Cascade lasers (QCL) [3-5]. Another approach to generate a THz FC, is down conversion/difference frequency mixing of an optical FC [6]. The common feature of these methods is a THz FC frequency grid defined by the effective length of the involved optical cavity. In the case of down conversion of mode-locked lasers, lines are typically spaced by 40 to 250 MHz, while THz QCLs deliver FCs with a line spacing of 6 to 30 GHz. Therefore, cavity modifications are necessary in order to change the number of comb lines, their spacing and the position of the comb in the THz range.

Near-infrared (NIR) electro-optic FCs (NIR EO FC) can be controlled electronically with a large degree of freedom [7] and their parameters are solely determined by the frequency stability and phase noise of the NIR light and the RF source [8]. Such electro-optic FCs are used e.g. for a self-heterodyne analysis of light sources [9,10], high resolution spectroscopy of atomic transitions [11] and high-sensitive interference spectroscopy [12].

In this work we show that optical EO FCs can be used to generate very flexible THz FCs. We realize this by employing difference frequency mixing in a semiconductor photomixing device [13]. This flexible THz opto-electronic FC light (TOFL) generation outperforms present THz comb sources in terms of the flexibility of the comb's spectral content. The key parameters of the presented source - comb linewidth and spectral brightness - demonstrate that the TOFL source meets the needs of chemical sensing under ambient conditions.

Our TOFL source is based on a NIR EO FC generated by phase modulation of a single-mode NIR laser which creates multiple optical sidebands forming a high-quality comb [7]. To generate the THz frequency comb, we mix the NIR EO FC with a second monochromatic NIR laser in a photoconductive THz antenna [13,14]. The resulting THz FC spectrum is given by the difference frequency spectrum of all involved NIR spectral lines. Ultrafast opto-electronic technology in the field of THz photonics has matured far enough that this conversion can be performed effectively [13,14] and high-resolution spectroscopy of gases is enabled [15]. Thus our new proposed opto-electronic source for flexible generation of THz FC is a merger of the two established technologies mentioned above - NIR EO FC generation and the generation of terahertz electromagnetic waves by photomixing of NIR light.

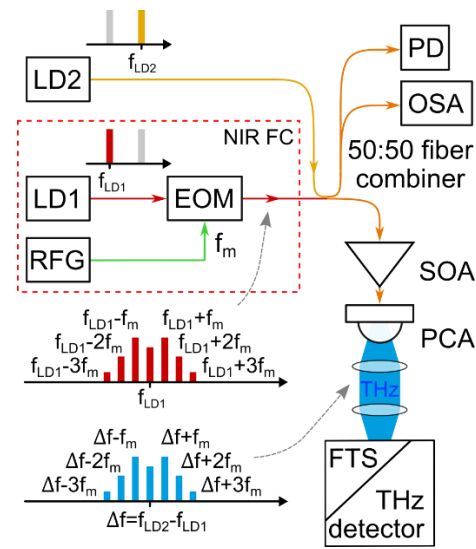

**Fig. 1.** Schematic drawing of the THz FC source setup: A near-infrared electro-optic frequency comb (NIR FC) source with a fixed-wavelength diode laser (LD1) and an electro-optic phase modulation block (EOM) driven by an RF generator (RFG); a tunable diode laser (LD2), a semiconductor optical amplifier (SOA), an optical spectrum analyser (OSA, Yokogawa AQ6319), a photo diode (PD); a photoconductive antenna (PCA) acting as NIR to THz opto-electronic converter. The generated THz beam is guided to a THz power detector or a Fourier Transform Spectrometer (FTS).

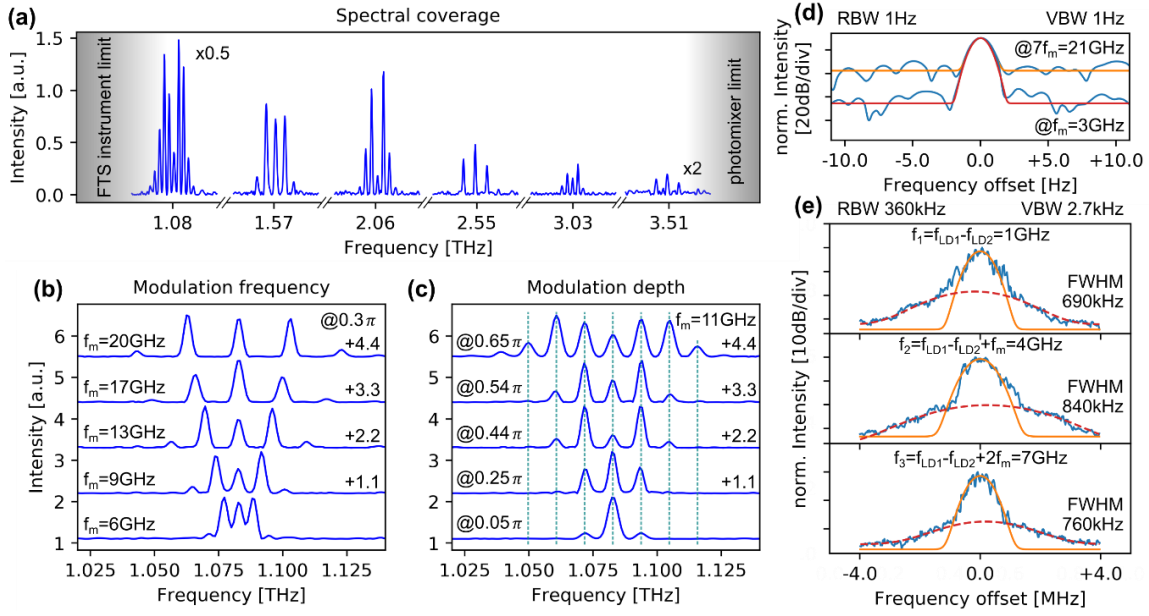

**Fig. 2.** Opto-electronically synthetized terahertz frequency comb spectra: (a) Spectral coverage achieved by detuning the NIR laser diodes LD2 (used phase modulation amplitudes and frequencies are listed in Table 1) as registered by a FTS with a spectral resolution of 2.7 GHz; (b,c) Electronic control of the comb design (line spacing and the power distribution); spectra are offset for clarity; dashed lines serve as a guide for the eye; (d) Intensity-normalised intermodal RF-beating spectrum of a NIR EO FC formed by phase modulation frequency of 3.0 GHz, 1.8s scanning time; (e) Optical beating of LD2 with the modulated output of LD1 measured by a fast photodiode (the RF-beating spectrum is fitted with two Gaussians to determine the beatnote linewidth on the background), 0.2s scanning time.

The principle of the TOFL source is shown in Fig. 1. A custom NIR EO FC is formed via phase modulation of a single-mode, fibre-coupled, stabilised NIR laser diode (LD1, Thorlabs SFL1550P, at 1555 nm) in a fibre-coupled electro-optic phase modulator block (EOM, iXBlue MPZ-LN-20). The modulator is driven by an RF source (RFG, Agilent N5173B) with an adjustable frequency, phase and output power. By controlling the RF power fed to the EOM, it is possible to define the achievable phase shift; thereby, the phase and amplitude of the modulation frequency uniquely determine the final THz comb shape.

Although more advanced modulation schemes may be employed [11,16], we demonstrate here that a simple harmonic RF driving of a single phase modulator already allows to cover a 100 GHz band with several frequency-equidistant modes. The addition of a tunable second NIR diode laser (LD2, New Focus TLB6370) enables to shift the original NIR EO FC into the THz frequency range. Both NIR beams are combined in a polarisation maintaining fibre and amplified by a semiconductor amplifier (SOA, Thorlabs BOA1004P) to 20 mW. This NIR radiation is fed into an opto-electronic converter ( PCA, Toptica EK000724) where the beating between the NIR spectral lines generates THz waves which emit THz radiation through the antenna structure. In this manner, the central frequency of a THz FC can be set by appropriately offsetting the wavelengths of the diode lasers LD1 and LD2.

In Fig. 2(a), we show several different THz FCs generated by our TOFL source. To demonstrate the comb design flexibility, THz FC spectra are generated at various combinations of phase modulation conditions (i.e. frequency $f_m$ and amplitude $\varphi_m$) while gradually tuning the emission frequency of LD2 by 20 nm (the used parameters are listed in Table 1). This tuning of diode laser LD2 allows the THz FC's central frequency to be shifted by ~2.5 THz.

**Table 1. Parameters for the THz FC generation shown in Fig. 2(a)**

| lasers detuning $f_{LD2} - f_{LD1}$ [THz] | modulation frequency $f_m$ [GHz] | modulation amplitude $\varphi_m$ [radians] |
|---|---|---|
| 1.083 | 9.0 | $0.66\pi$ |
| 1.573 | 17.0 | $0.38\pi$ |
| 2.062 | 11.0 | $0.47\pi$ |
| 2.547 | 22.0 | $0.19\pi$ |
| 3.030 | 10.0 | $0.55\pi$ |
| 3.507 | 21.0 | $0.44\pi$ |

The generated THz combs are analysed by a Fourier transform spectrometer (FTS, Bruker Vertex 80). They cover a range of 1.1-3.5 THz that is limited on the low-frequency side by the cut-off wavelength of the FTS. The upper-frequency limit of the demonstrated THz FC is set by the frequency response of the THz photoconductive antenna [14]. Switching between different THz FCs can be done almost instantly; it is mainly limited by the time required for the wavelength re-adjustment of laser diode LD2. The total power of the generated THz FC was measured by a Golay cell (Tydex Ltd.) with a responsivity of 70 kV/W and lies in the microwatts power range.

The all-electronic, flexible control of TOFL spectra is further illustrated for a fixed frequency offset of the laser diodes (LD1 and LD2) frequency, as shown in Figs. 2(b) and 2(c). The frequency spacing and power distribution between the THz comb lines are simply set by the phase modulation parameters. Besides the spectral coverage and design flexibility, the linewidth (i.e. the integrated phase noise) of the generated THz FC lines is of significant importance for the practical use of a TOFL source. According to the theory of driven oscillators, the linewidth of the oscillator frequency corresponds to the phase noise of the driving force; hence, the linewidth of THz emission should be given by that of the beating NIR frequencies. The used single-mode laser sources (LD1 and LD2) have linewidths of < 200 kHz. The NIR EO FC derived from LD1 by a phase modulation is measured and features an intermodal beating linewidth of about 2 Hz (Fig. 2(d)). The comb lines only marginally deteriorate with increasing line index as the linewidths of the intermodal beating of the lines separated by $1 \times f_m$ and $7 \times f_m$ demonstrate. Thus, the line broadening that comes with the frequency multiplication via phase modulation is much smaller than the system phase noise. This proves that the phase fluctuations of both the laser diode LD1 light and the RF source are coherently transferred to all comb lines. This behaviour guarantees that every THz comb line resulting from the photomixing of the NIR EO FC with the light from the second diode laser will feature similar phase noise characteristics.

To estimate the linewidth of the generated THz FC lines themselves, we applied two different procedures. First, the emission frequencies of the two laser diodes were offset by 1 GHz, and the phase modulation frequency $f_m$ was set to 3 GHz. Therefore, all resulting optical beatings (i.e. the sub-THz FC lines) fall within the bandwidth of 15 GHz of an ultrafast InGaAs photodetector (Thorlabs DET08CFC/M) and are measured by an RF spectrum analyser (Agilent CXA N9000A). We found that the three neighbouring comb lines exhibit similar linewidths below 1 MHz (Fig. 2(e)), which can be considered as a conservative estimate for the lower bound for the THz comb linewidths, although not the THz but the optical beatnote is measured. This value reflects the phase noise contributions from the diode lasers, the RF generator and the used photodetector signal chain.

The electronic noise and frequency response of the opto-electronic converter PCA, as well as the mechanical vibrations of the THz emitter/detector opto-mechanics, can contribute to an excessive phase noise. Therefore, in the second procedure, we directly studied the linewidth of the THz comb lines. Since the typical resolution of an FTS instrument is on the order of several GHz, it cannot resolve the true THz frequency linewidth. Hence, we determined the linewidth by tuning one comb line frequency through a molecular absorption region. We have chosen the far-infrared active rotational transition of the ammonia ($NH_3$) molecule at 572.498 GHz (19.09648 cm$^{-1}$). Figure 3 shows the results of the measured transmission through a 70-mm-long stainless-steel home-made gas cell with quartz windows filled with $NH_3$ (3,5N purity) at a pressure between 6 and 20 mbar and a temperature of 296 K. We used a THz comb line corresponding to a frequency of $f_{LD2} - f_{LD1} + f_m$; the comb line was scanned through the absorption line of $NH_3$ by sweeping the modulation frequency $f_m$ of the phase modulator (i.e. varying the comb spacing) with a constant modulation amplitude $\varphi_m$, while keeping the other comb lines outside of the scanned frequency window. No spectrometer is used. The optical power of all the comb lines falls on a single THz detector and the experiment relies on the assumption that only one comb line is in resonance with a molecular transition. As expected, the transition linewidth broadens with increasing $NH_3$ gas pressure. All experimentally determined linewidths follow the broadening values obtained from the HITRAN database [16] with somewhat increasing error margins for higher pressure values. We obtained a pressure-induced broadening of 30.69 ± 2.65 MHz/mbar, while the calculations based on the HITRAN database yields a value of 29.65 MHz/mbar which is lying within the error margins of the experimental data.

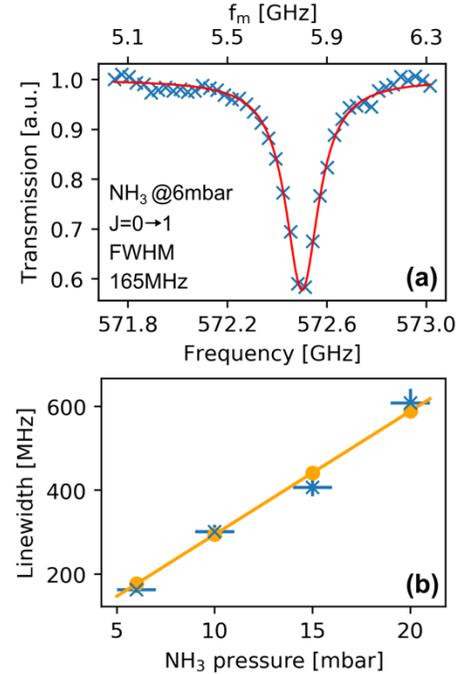

**Fig. 3.** $NH_3$ molecular rotational transition at 572.498 GHz (19.09648 cm$^{-1}$): (a) transmission (blue crosses) measured at a pressure of 6 mbar and a temperature of 296 K (red line - Lorentzian fit used for linewidth determination). (b) Measured pressure broadening of the absorption line (blue crosses) compared to the linewidth values (yellow dots) obtained from the HITRAN database [17].

The experimentally obtained spectral profile of the absorption line, as shown in Fig 3(a), is a result of the convolution of the THz comb line with the collision-broadened rotational transition and their values indicate that the linewidth of the THz FC teeth must be < 10 MHz. This upper bound for the linewidth of the THz comb line is a very good value for a system consisting of two independent laser sources that use only a temperature stabilisation of the cavity. We assign the observed multi-MHz linewidth of the THz FC teeth to a residual relative frequency drift of the semiconductor diode lasers during scanning through the frequency window of the molecular transition. Indeed, we observed a maximum broadening of the beatnote of 10 MHz during the data acquisition performed by the RF signal analyser in a minute time window.

The current implementation allows a continuous spectral coverage from 0.5 THz to 3.5 THz, which is an order of magnitude larger than other EO based techniques [18, 19] or utilizing the non-linearity of a THz photomixer for THz comb generation [20]. A comb linespacing up to 22 GHz can be reached as well as a maximum number of about 11 comb lines. The number of lines is determined by the achievable modulation amplitude limit of $2/3\,\pi$, which can be improved by using more advanced modulation schemes [11,16] and would lead to a substantially widened parameter space for the THz comb design. Our simple approach to generate THz FCs compares very well with QCL THz FCs in terms of the number of comb lines within 10 dB intensity variation [3-5,21-23].

The linewidths of our TOFL source lines are already better than 10 MHz and thus very attractive for many sensing applications. A deep sub-MHz frequency precision could be reached by locking the diode lasers to a reference absorption line [24] or to a NIR FC [25,26], by electro-optical frequency shifting [27], while a mode filtering approach guarantees a Hz level stability of the generated THz frequency [18]. The linewidth of 10 MHz also sets the limit for the frequency resolution of the current version of TOFL and is comparable to that of QCL combs [22,23]. Here we emphasize the superiority of THz generation schemes using NIR lasers which are frequency stabilized to NIR FCs [18,25,26,28]. However, their ~ 1 Hz frequency precision comes with large, complex setups and is not required for the majority of potential THz sensing applications.

The power generated per comb line outperforms that of THz FC sources based on NIR ML lasers; it is comparable to other CW-THz generation systems [25, 26], and is about 10 dB lower compared to THz QCL FC sources [3-5,21-23].

The standard fibre-optic components used in our TOFL source lead to a robust setup that provides straightforward handling of THz FC generation with guaranteed frequency flexibility. In the long term, since our TOFL source utilises the principles and technology of optical fibre-based communication, the optical fibre infrastructure for the distribution of frequency standards derived from atomic clocks, would allow the local generation of precise custom THz spectra on demand.

**Funding.** The Austrian Science Fund (FWF DK CoQuS W1210 & FWF DiPQCL P30709-N27).

**Acknowledgment.** The Authors acknowledge Thomas Müller (TU Wien) for providing the diode lasers, Michael Feiginov (TU Wien) for helping with room temperature THz detection and TU Wien Bibliothek for financial support for the proofreading.

**Disclosures.** The authors declare no conflicts of interest.

**Data availability.** The data presented in this paper are available from the corresponding authors upon reasonable request.